# A Metric for Modelling and Measuring Complex Behavioural Systems

Kieran Greer

Distributed Computing Systems, Belfast, UK.

http://distributedcomputingsystems.co.uk

Version 1.4

*Abstract:* This paper[1] describes a metric for measuring the success of a complex system composed of agents performing autonomous behaviours. Because of the difficulty in evaluating such systems, this metric will help to give an initial indication as to how suitable the agents would be for solving the problem. The system is modelled as a script, or behavioural ontology, with a number of variables to represent each of the behaviour attributes. The set of equations can be used both for modeling and as part of the simulation evaluation. Behaviours can be nested, allowing for compound behaviours of arbitrary complexity to be built. There is also the capability for including rules or decision making into the script. The paper also gives some test examples to show how the metric might be used.

**Keywords:** Autonomous Behaviour, Complex System, Modelling Script, Metric.

---

# 1    Introduction

This paper describes a set of equations that can be used to estimate the success of a group of agents that are required to solve a particular problem. The context of these agents is an autonomous environment, where the agents can perform a number of different, but relatively simple acts, without the help of a guiding system. In this environment it is difficult to control or fully predict the outcome of the agent activities and so it would be helpful if a metric could give an initial guidance or guess as to how successful the agents are likely to be. The set of equations, or framework, can therefore be used in two different ways. At the modelling stage, when you are guessing what values each agent should have, you can use the equations with fixed values. This can produce scores of the likelihood of success, using those agents specifically. You would therefore model the problem, defining attribute values and then run a simulation with those values. If the simulation is not satisfactory, you are able to change the model set of values more easily, as you would know what value relates to what. To evaluate a dynamic problem or run a simulation, you can still use the equation framework and plug in dynamic formulae, for example, whose input values then change as part of the simulation.

The papers [1] and [2] describe in some detail the need for metrics for testing complex systems. The problem is that if a system is set in motion that contains a large number of these agents and is expected to perform certain tasks, then it would be helpful to know with some confidence that the system will in fact be able to carry out the required tasks. This would save time modelling the problem, if the agents are discovered to be too simple and also with evaluating the result of any simulation. If something goes wrong with the process, then it might be difficult to identify the source of the problem – is it agent1 and behaviour *x* or agent2 and behaviour *y*? So metrics would be helpful to allow a user to guess or estimate as to how successful the system is likely to be. Also, if the evaluation suggests that the system will not work, it provides a way to change parameters at the design stage and re-evaluate, before implementing and running a system in a real or test environment.

These systems come under the umbrella term of Complex Adaptive Systems [3] and are typically represented by a network of agents, acting in parallel and reacting to changes in their environment, to realise some greater goal. Typical features of CAS include decentralised control, autonomous dynamic behaviour, and chaotic or stochastic results. This is also closely related to the modelling of simple biological societies in complex systems, for example, the stigmergic termite societies [4]. There are several examples of the bio-inspired stigmergic behaviour being applied in complex systems, usually for self-organising or optimisation problems. With stigmergy, the relatively simple agents cooperate without any central guiding system, to perform more complex tasks.

The metric that will be defined is based on the characteristics of stigmergic systems [5]. These characteristics however, also appear to be a general kind of definition for simple biological agents. The metric would also be suitable for more complex modelling, because it provides a framework into which more complex functions can be added. It is based on specifying behaviours with related attribute values, where these behaviours can be nested to realise compound behaviours of arbitrary complexity. When modelling the system, the actions of the agents are defined by these behaviours, such as 'move in a particular direction', or 'lift an object'. Agent 'types' are defined by what particular behaviour types they can perform and then a number of agent instances are allocated to each agent type.

The outline of the rest of the paper is as follows: section 2 describes the essential behavioural components of stigmergic behaviour that the metric is based on. Section 3 describes some related work. Section 4 describes the new metric and related attributes. This includes the equations used to evaluate the problem success. Section 5 shows how the metric and attributes can be used to model the behavioural characteristics of an autonomous system. Section 6 describes some test cases based on known problems, while section 7 gives some conclusions on the work.

## 2  Stigmergic Behaviour

There are a number of papers that describe the behaviour that can emerge through stigmergic principles, for example, [5] and [6]. The authors of [5] also define the main components of collective behaviour as coordination, cooperation, deliberation and collaboration.

- *Coordination* – is the appropriate organisation in space and time of the tasks required to solve a specific problem.
- *Cooperation* – occurs when individuals achieve together a task that could not be done by a single one.
- *Deliberation* – refers to mechanisms that occur when a colony faces several opportunities. These mechanisms result in a collective choice for at least one of the opportunities.
- *Collaboration* – means that different activities are performed simultaneously by groups of specialised individuals.

They note that these are not mutually exclusive, but rather contribute together to accomplish the various collective tasks of the colony. They also note that a swarm intelligence (for example, stigmergy) community also needs to be able to adapt their behaviours on an individual level, due to changes in the environment, or the group composition itself. They conclude by explaining that these four characteristics cover aspects of space and time (coordination), social structures (collaboration), decision making (deliberation), and team work (cooperation).

## 3  Related Work

This section notes some points made by other researchers, outlining the need for such metrics. The papers [1] and [2] describe in some detail the need for metrics for testing complex systems. Nguyen et. al. [2] makes the important point:

> 'Testing traditional software systems, which have reactive (or input-output) style behaviour, is known to be non-trivial, but testing autonomous agents is even more challenging, because they have their own reasons for engaging in proactive behaviour that might differ from a user's concrete expectations, yet still appropriate; the same test input can give different results for different executions.'

This states that even if you can model the environment correctly, you can still not evaluate with certainty how the agent behaviours will evolve over time or what they will finally produce, because the autonomy in the system can produce different results for the same set of initial criteria. To test the system for this sort of thing would require a simulation environment that allows the agents to play out their behaviours and their results to be measured during the simulation. This paper is concerned more with providing an initial idea of whether the agents to be used have the actual capacity to be able to solve the problem. For example, if the system is required to lift a one tonne block, but there are only two agents that can lift only 100 kilograms each, then you can determine that the system will fail. There is no need to play out a simulation, because it will simply be impossible. If however there are 100 agents, then it might be possible. The scripting language would also allow a simulation program to run based on it however, and allow the related agents to effectively evaluate their own actions.

The paper [7] also describes the need for test-beds and metrics for testing complex systems. They describe some of the test environments for agent-based systems that were available at the time. The TILEWORLD [8] environment will be used in section 6 as part of a test example. They argue [7] that test beds provide metrics for comparing competing systems, but they can also be used to highlight interesting aspects of system performance. This evaluation is realised however, only if the researcher can adequately explain why his or her system behaves the way that it does. Understanding a system's behaviour on a benchmark task requires a model of the task. Without this model it is difficult to determine what has been accomplished. 'We risk finding ourselves in the position of knowing simply that our system produced the successful behaviour - passing the benchmark.' Models are also important when designing benchmarks to be failed, so that critical factors or elements in the system can be recognised for that situation as well. Allowing multiple agents to act in the world also introduces problems such as coordination and communication, and how the effects of simultaneous actions differ from the effects of those actions performed serially. The modelling of more complex behaviours eventually leads to game-playing programs with autonomous characters. One example of this is given in [9] and this test-bed will be used in section 6.2 to outline one important aspect of the model. In [6] it is noted that it would be helpful in modelling these kinds of systems if the following questions could be answered:

'How can we define "methodologies" to program ant algorithms? How do we define "artificial ants"? How complex should they be? Should they all be identical? What basic capabilities should they be given? Should they be able to learn? Should they be purely reactive? How local should their environment knowledge be? Should they be able to communicate directly? If yes, what type of information should they communicate? What is also missing, similarly to what happens with many other adaptive systems, is a theory that allows to predict the system behaviour as a function of its parameters and of the characteristics of the application domain.'

The metric described in this paper is closely associated with these questions and can be used to help with the modelling and prediction problems that are written about. The relatively new 'formal' subject of Agent-Based modeling would also be able to use this sort of behaviour script. The paper [10] notes that behaviour is often missing when modeling complex systems, but goes into all aspects of agent-based modeling in more detail.

## 4  Behaviour Metric

Modelling the agent behaviour has led to the following list of attributes being considered as part of the model. The variables are as follows:

1. *Individual agent characteristics*: These are characteristics that relate to an agent as an individual. If an agent is performing the specified behaviour, then when acting by itself as an individual it will have the following characteristics:

1.1. *Ability*: this defines how well the behaviour is able to execute the required action. A high value means that it is very good at the action and a low value means that it is very bad at the action.

1.2. *Flexibility*: this defines how well an agent performing this behaviour can adapt or change to a different behaviour if the situation requires it to. This could be an internal decision or a reaction to an external event. This can be seen as the ability to make that decision before any other action is taken. The collective capabilities of coordination and collaboration described next can then be seen as the ability to be flexible after that decision has been taken.

2. *Collaborative agent characteristics*: These are characteristics that relate to the agent working in a team environment. If an agent is performing the specified behaviour, then it will be able to interact with other agents with the following performance level:

2.1. *Coordination*: this defines how well the agent can coordinate its actions with those of other agents, if certain discrete actions need to be performed together. This is related to flexibility, but this variable measures the group aspect of the attribute after the decision to be flexible has already been taken.

2.2. *Cooperation*: this defines how well an agent performing an action can cooperate with other agents also involved in that action. For example, if several agents need to lift an object together to move it, how well can they work together to do this?

2.3. *Communication*: this defines how well the agents can communicate with each other. This is defined as an input signal and an output signal for each behaviour type. As different behaviours could occur under different circumstances and an entity can be made from more than one behaviour, the communication capabilities could be different for the same entity. For example, if all agents are lifting a block together, they need to communicate in close proximity, to describe the direction of movement. If the agents are looking for the block, they may need to communicate over long distances and relate a different kind of information.

The metric is quite well balanced, with approximately half of the evaluation going to the individual capabilities and half going to the group capabilities. This also maps across to the principles of stigmergic behaviour described in section 2. Deliberation maps to the agent's intelligence, while the social, team work and relative space/time co-ordinates map to the collaborative attributes. Note that if different agents perform the same behaviour but at different levels of success, this would be defined by different types of the same behaviour. Alternatively, if there are underlying equations that calculate each of the attribute values, these could dynamically produce different behaviour values for the same behaviour type. This will be written about more in section 5, but the static values are more likely to be used during modeling and the dynamic values as part of a simulation. With these main variables, it is possible to define a set of equations that can be used to measure how likely it is that a certain problem can be solved by the agents that are available. These are described next.

**4.1 Equation Groups**

These equations are built from the definitions of the previous section. Each variable can be assigned a value based on the overall design goals and available resources, but the equations do not define how this should be done. If taken as a kind of framework, the individual values could be general evaluations (poor, average, good, etc.), or they could come from other, possibly non-linear equations, providing more complex or accurate values. The intention is that these equations and the related modelling script, cover the main aspects of the behavioural problem and are flexible enough to be useful for a wide variety of problems. This will also provide a consistent model over the whole problem domain. The equations can be divided into those that evaluate the problem and those that evaluate the agents that are used to solve the problem. All evaluations are normalised to be in the range 0 to 1.

**4.2 Problem Related**

There are essentially two considerations here – how complex the problem is and how likely it is that it can be solved. The Problem Success Likelihood (PSL) is the final problem evaluation and estimates how well the problem can be solved. This is not related to the known statistical Likelihood equations, but tries to measure a similar sort of concept. The top part of the PSL equation, shown in equation 1, evaluates the average agent complexity ($EC_s$), which is a combination of all of its attributes. This is measured for all of the behaviour type instances ($B_s$) that are part of the problem behaviour set (PBS). This can be no larger than the optimal problem complexity (PC) value of 1.0. The problem complexity is a factor of how intelligent the agents need to be to solve it, including their collective or interaction capabilities. Because the evaluations are all normalised, the maximum value that the problem complexity can be is 1.0. If all behaviours are perfect, they will also only sum to 1 as well and so to normalise that does not require any further divisions. The variable has been kept just as a reminder. The problem success likelihood, can therefore be defined as follows:

$$PSL = \frac{(\sum_{s=1}^{n} EC_s)/n}{PC} \quad \forall\ Bs \in PBS \qquad \text{Eq. 1}$$

### 4.2.1 Individual Parameters

The following equations that are used in equation 1, can be used to measure the behaviour's individual capabilities.

$$EC_s = \frac{I_s + COL_s}{2} \qquad \text{Eq. 2}$$

$$I_s = \frac{BA_s + BF_s}{2} \qquad \text{Eq. 3}$$

The agent or entity complexity ($EC_s$) for behaviour $s$ is a factor of its ability to perform the related behaviour attributes of intelligence and collective capabilities. The agent intelligence ($I_s$) is a factor of its ability ($BA_s$) and flexibility ($BF_s$) capabilities for the specified behaviour.

### 4.2.2 Team Work

The following equations can be used to measure the behaviour's collective or team work capabilities ($COL_s$).

$$COL_s = \frac{(COR_s + COP_s + COM_s)}{3} \qquad \text{Eq. 4}$$

$$COM_s = \frac{SI_s + SO_s}{2} \qquad \text{Eq. 5}$$

The collective capabilities of the agent performing the behaviour are its ability to cooperate with other agents ($COP_s$), coordinate its actions with them ($COR_s$) and also communicate this ($COM_s$), normalised. The communication capabilities of the agent for the behaviour *s* include its ability to send a signal to another agent ($SO_s$) and also its ability to receive a signal from another agent ($SI_s$).

## 5  Modelling the Problem

Using the metric of section 4.2, it is possible to specify a problem with all of the related agents and actions that are part of the solution space. The modelling is based around the behaviour types that are used to solve the problem, where the same type definitions can be used to model the problem and also to simulate its execution. The agents are defined by agent types, where an agent type can perform a particular set of behaviours. So if the same behaviour type is to be performed at different levels of success; for static values this would require different behaviour definitions, or for dynamic ones the value can change. The problem itself is the same behaviour set, where the required tasks are specified by the behaviours that can perform each task. For example, a problem of moving a block could be defined as:

- Task – Move Block to Destination
    - Behaviour 1 – Lift Block.
        - Number of agents required 2.
    - Behaviour 2 – Move Block.
        - Number of agents required 2.

These behaviours can be represented by the same behaviour types that would represent the agent types. However, the 'problem specification' behaviours can be assumed to work perfectly, that is, all of the variable values can be set to 1.0. The agents involved will then perform the 'same' behaviour types, but the behaviour definitions will have different attribute values depending on how well they can be executed. There are then a number of agent instances that are defined as being of a particular agent type. If a task's goal requires 2 agents for example, the two individual agent scores are summed together and divided by the agent number, to produce an average value.

The specification also allows behaviours to be made more complex, by adding sub-behaviours. For example, to move the block, the agent must move in some direction, where the behaviour definition could look like:

- Behaviour – Move Block.
    - Sub-behaviour – Move North.
    - Sub-behaviour – Move South.
    - Sub-behaviour – Move East.
    - Sub-behaviour – Move West.

These can either be required or alternative sub-behaviours. For example, to move the block, the movement is in one of the specified directions, not all of them at the same time. When providing a score for the 'Move Block' behaviour, the scores for the sub-behaviours are also included. If there are alternative sub-behaviours however, then they can have different attribute values and so this leads to upper and lower bounds on the compulsory behaviour score. The compulsory score is the score for all behaviours that are required for a single action. The lower bound is then this score plus the lowest score for any of the alternative behaviours. The upper bound is then this score plus the highest score for any of the alternative behaviours. For example, to move the block, the compulsory behaviour is the 'Move Block' behaviour. Also required however is one of the alternative behaviours, required for moving in a particular direction. So the bounds values would consider the compulsory behaviour plus one of the alternative behaviours. If the agent has difficulty in moving in certain directions, this could produce a range estimate of how well the behaviour is likely to fulfil the task in general, such as:

    Compulsory evaluation: 0.75.
    Lower bound: 0.71
    Upper bound: 0.79.

Another factor is whether the agent can work alone or requires the help of other agents. If the action is for a single agent, then the collaborative capabilities are not considered, but if it is as part of a team, then they are. If an attribute is not required, its value can be set to 1.0, when it will not then affect the overall score for the behaviour.

When defining a behaviour, the model therefore allows for any level of complexity. This means that a behaviour could be seen to be simply a single action, or even a single mental act, such as making a decision. This leads to the idea of programming rules into the script, where one behaviour that is a mental decision would support the parent behaviour, while another one might prevent it. This is also easily added with an additional attribute to the sub-behaviour definition that defines if the sub-behaviour is positive or negative, with respect to the parent behaviour. If it is positive then its evaluation is added and averaged with the parent behaviour as normal, but if it is negative, then the negative equivalent of its evaluation is added instead.

Some points to note are as follows: The first point is the difference between the evaluation metric and a simulation scenario. The evaluation metric can estimate how well the system will work as a whole, which is a measure of how well each behaviour can be performed, including doing the negative actions 'well'. In that case, all of the behaviour scores should be added together positively, as it is just the overall level of 'ability' that should be measured. The modelling scores are probably also an upper and lower bound on the problem execution result, for if all agents cooperate perfectly with each other, they will achieve the best problem likelihood estimate only. They would have to choose the worst behaviour or act each time to then achieve the lower bound score. In a simulation environment, one particular action might depend on a previous decision that can be either positive or negative. In that case, the evaluation for the selected sub-behaviour can either be added or subtracted from the parent evaluation. Also, as the sub-behaviours can also have bounds, a range of values is possible for negative or blocking actions as well. If the negative behaviour can be evaluated with certainty and must also block the parent action, then provision could be made for that relatively easily.

# 6   Test Example

This section will describe two different test examples. These only simulate one example situation in the test-bed problem, but show how the metric and script would be useful for solving these types of problem.

## 6.1  Modelling a Real Test Problem

A test example can be based on the classic-style TILEWORLD problem. This environment is made up of a grid-like world, where agents can move to one of the cells that is adjacent to the one that it is on. The cells can also contain obstacles, tiles or holes. The aim is for the agent to move round the world, avoiding the obstacles and filling the holes with tiles. The agent can only move a tile if it is currently occupying a cell that is directly adjacent to the tile. In this version of the game, some of the tiles require more than one agent to be moved and so the agents will need to cooperate with each other to do that. The following behaviours and conditions can be defined:

- *Movement behaviour* – so that an agent can move about its environment and also avoid obstacles.
- *Lift / Carry behaviour* – so that an agent can move a tile to a new location.
- *Intelligence* – can help the agent decide where to move, or how to plan its actions. This can also help to determine what hole should be filled by the tile, for example.
- *Flexibility* – for example, if an agent is at one tile and is asked by several others to help pick up another tile, then it should perform the multi-agent task first as it is a more rare occurrence.
- *Coordination* – so that an agent can determine when to pick up one tile with other agents before moving somewhere else. This is a group decision as opposed to the individual agent decision.

- *Cooperation* – so that the agent can pick up a tile with other agents when required and they can all move it in the same direction.
- *Communication* – this could define the distance between which agents can receive requests and possibly the complexity of those requests.

With these parameters, a description of the problem can be written in a scripting language such as XML, as shown in Figure 1. This can then be read and processed in an automated way. The described problem has been simplified into the task of moving the tiles into the holes. The behaviours can then be described in a similar way, for example, as shown in Figure 2. This example shows that the behaviour to move a tile also requires the behaviours to move in one of the four directions, but these are alternative options where only one would be executed as part of the next action. These sub-behaviours can then be defined later in the script. See Appendix A, for example.

```
<Problem_Complexity>
    <Problem_Task Name="Move Tile into Hole">
        <Problem_Behaviour Type="Lift Tile">
            <Entity_Number>2</Entity_Number>
        </Problem_Behaviour>
        <Problem_Behaviour Type="Move Tile">
            <Entity_Number>2</Entity_Number>
        </Problem_Behaviour>
    </Problem_Task>
</Problem_Complexity>
```

```
<Behaviour Type="Move Tile">
    <Ability>1.0</Ability>
    <Flexibility>1.0</Flexibility>
    <Collective>
        <Coordination>0.5</Coordination>
        <Cooperation>0.5</Cooperation>
        <Communication>
            <Signal_In>0.5</Signal_In>
            <Signal_Out>0.5</Signal_Out>
        </Communication>
    </Collective>
    <Requires>
        <Behaviour_Type AndOr="Or">Move North</Behaviour_Type>
        <Behaviour_Type AndOr="Or">Move South</Behaviour_Type>
        <Behaviour_Type AndOr="Or">Move East</Behaviour_Type>
        <Behaviour_Type AndOr="Or">Move West</Behaviour_Type>
    </Requires>
</Behaviour>
```

Figure 1. Problem task description          Figure 2. Example description of a behaviour

## 6.2 Test Results

The figures of Table 1 give an example of what the behaviour variables and values might look like and how changing these could change the success of solving the problem. If only one agent is required to lift a tile, then the collective capabilities are not required and the estimated success values are shown in the '1 Agent' column of the table. This is because the ability and flexibility values are all set to 1.0 in the behaviour descriptions. Figure 1 shows that two agents are in fact required for each behaviour in the task. This means that the agents will need to cooperate and so their collective capabilities are required. If all of the movement behaviour attribute values are the same as the 'Move Tile' attribute values shown in Figure 2, then the estimated success values are shown in the '2 Agent' column of the table. This is calculated from a perfect individual score of 1.0 and a collaborative score of 0.5, averaged for each agent. If however, some of the attribute values are made different, then the evaluation can be given upper and lower bounds. For example, if we take two of the movement behaviours - 'Move North' and 'Move South' - and reduce their collective values to be 0.25 instead of 0.5, the estimated success values are shown in the '2 Agent + Worse Behaviours' column of the table.

Table 1. Different evaluations for different agent combinations

|  | 1 Agent | 2 Agents | 2 Agents + Worse Behaviours |
|---|---|---|---|
| **Evaluation** | 1 | 0.75 | 0.75 |
| **Upper Bound** | 1 | 0.75 | 0.75 |
| **Lower Bound** | 1 | 0.75 | 0.74 |

If either of the worse movement directions is taken, then the total behaviour value is not as good and so there is a smaller lower bound. For testing or modeling, it is then possible to change the values incrementally, to calculate what this might do to the overall success. For example, if we increment, or improve, the coordination and cooperation variables of 'Move North' by 0.05 for 10 iterations, the success value will change as shown in Table 2. Note that the other worse movement behaviour (Move South) is keeping the lower bound as it is. If this is then also incremented by the same amount, the values will change to those shown in Table 3.

Table 2. Bounds produced after incrementing the variables of one behaviour

| Iteration | Evaluation | Upper Bound | Lower Bound |
|---|---|---|---|
| 1 | 0.75 | 0.75 | 0.74 |
| 2 | 0.75 | 0.75 | 0.74 |
| 3 | 0.75 | 0.75 | 0.74 |
| 4 | 0.75 | 0.75 | 0.74 |
| 5 | 0.75 | 0.75 | 0.74 |
| 6 | 0.75 | 0.75 | 0.74 |
| 7 | 0.75 | 0.75 | 0.74 |
| 8 | 0.75 | 0.751 | 0.74 |
| 9 | 0.75 | 0.752 | 0.74 |
| 10 | 0.75 | 0.753 | 0.74 |

Table 3. Bounds produced after incrementing the variables of both behaviours

| Iteration | Evaluation | Upper Bound | Lower Bound |
|---|---|---|---|
| 1 | 0.75 | 0.75 | 0.741 |
| 2 | 0.75 | 0.75 | 0.742 |
| 3 | 0.75 | 0.75 | 0.744 |
| 4 | 0.75 | 0.75 | 0.745 |
| 5 | 0.75 | 0.75 | 0.747 |
| 6 | 0.75 | 0.75 | 0.748 |
| 7 | 0.75 | 0.75 | 0.749 |
| 8 | 0.75 | 0.751 | 0.75 |
| 9 | 0.75 | 0.752 | 0.75 |
| 10 | 0.75 | 0.753 | 0.75 |

These two changes show that the upper bound eventually increases because these two behaviours ('Move North' and 'Move South') start to produce a larger value after the increments. In the first example the second behaviour keeps the lower bound down, but in the second example this is able to increase as well. It then stops at 0.75 because the originally better behaviours that do not change their values, will only produce this value. These equations are very linear and changes to the success evaluations will take a linear path, but this is also because the increments or changes themselves are constant.

### 6.3 Test Example of Rule or Decision Making

A suitable example of the decision-making capabilities would be the modelling of the test-bed game-playing environment described in [9]. This describes a strategy game called Legion II with two sides – Romans and Barbarians. The Romans have to patrol the playing area and protect the settlements from attack by the Barbarians. One of the rules in the game is that a Roman garrison may not leave a city if there are Barbarians as close as, or closer, to the city than them. If they then leave the city the Barbarians would be able to attack it and so they must go back to defend the city. This behaviour can be modelled by the script of Appendix A. This is a full script, but it only models this one task. In this example, the Romans can detect with certainty if there are Barbarians present, but if the Barbarians are not present, or possibly hiding, then they can only detect this with 50% ability. Evaluating the metric leads to the value:

Compulsory evaluation: 1.
Upper bound: 1.
Lower bound: 0.938.

If evaluating the behaviour as part of a simulation however, the decision to roam the countryside leads to the following result:

- Barbarians are present: roam countryside decision evaluation is 0.0.
- Barbarians are not present: roam countryside decision evaluation is 0.75.

If however the garrison finds the x-ray camera and can see the Barbarians even when hiding, this will change the 'Barbarians Not Close' variable values to be all 1.0. Evaluation of the metric then gives the values of:

Compulsory evaluation: 1.
Upper bound: 1.
Lower bound: 1.

If evaluating the behaviour as part of a simulation, the decision to roam the countryside now leads to the following result:

- Barbarians are present: roam countryside decision evaluation is 0.0.
- Barbarians are not present: roam countryside decision evaluation is 1.0.

And so with certainty, if the Barbarians cannot be detected, the garrison is free to roam the countryside.

## 7  Conclusions

This paper presents a metric and behavioural script that can be used to model complex systems exhibiting autonomous behaviours. The model tries to include the features of individual capabilities, collective capabilities and communication. The terms are rather general and so they could be flexibly used but do require arbitrary input, in the sense that it is not pre-defined. This input could be manual, but there is also the possibility of plugging in more complex or non-linear equations to calculate the values of each attribute. This would allow agents to perform the same behaviour at different levels of success, but still maintain the same overall structure. Other interesting aspects of the model might be the nesting of sub-behaviours, allowing more complex behaviour patterns to be built up. The behaviour can then be defined from the lowest level of a single mental act to a much more complex physical action. The idea of rules can also be programmed into the script, with certain decisions preventing an action while other ones would allow it to happen. Another critical aspect of this is the ability to provide lower and upper bounds on the behaviour success, which would make it slightly fuzzy, with the additional information that this provides.

The metric can give a guess to the likely success of solving the problem, but with a large number of variables, if the success is not good, then it is still difficult to decide what variable should be changed. Modelling round the behaviour and providing a consistent model means that changes can be made to each behaviour variable in a systematic way and the resulting overall change will reflect all agents of that type. It would be possible to find the most important or critical variables for a problem – the ones that made the most impact when changed, by testing and measuring each variable in turn. This would help the designer decide what changes would need to be made if the system was estimated to be unsatisfactory.

The inclusion of a flexibility attribute suggests something more along the lines of an animal behaviour. With increasing interest in the autonomous supervision and modelling of humans in the Medical domain, this sort of behavioural model might be useful in that environment as well, or even for modelling behaviours in AI robotics. The most important factor missing from the metric at the moment is probably time. This would be difficult to include at the modeling stage however and would probably also be linked more closely to the numbers of agents available at any moment in time. When modelling, the metric assumes that all agents are always available and so only the minimum requirement needs to be covered. In a simulation environment, time might be measured more accurately.

**APPENDIX A - Legion II Game-Playing Script Example**

```xml
<Problem_Spec>
    <Problem>
        <Problem_Complexity>
            <Problem_Task Name="Roam Countryside">
                <Problem_Behaviour Type="Leave Settlement">
                    <Entity_Number>1</Entity_Number>
                </Problem_Behaviour>
                <Problem_Behaviour Type="Move about Countryside">
                    <Entity_Number>1</Entity_Number>
                </Problem_Behaviour>
            </Problem_Task>
        </Problem_Complexity>
        <Problem_Behaviour_Set>
            <Behaviour Type="Leave Settlement"/>
            <Behaviour Type="Move About Countryside"/>
        </Problem_Behaviour_Set>
        <Problem_Entities>
            <Entity Type="Garrison">
                <Entity_Number>3</Entity_Number>
            </Entity>
        </Problem_Entities>
    </Problem>
    <Entities>
        <Entity Name="Garrison 1" Type="Garrison"/>
        <Entity Name="Garrison 2" Type="Garrison"/>
    </Entities>
    <Entitiy_Types>
        <Entity_Type Name="Garrison">
            <Entity_Behaviours>
                <Behaviour_Type>Leave Settlement</Behaviour_Type>
                <Behaviour_Type>Move about Countryside</Behaviour_Type>
                <Behaviour_Type>Barbarians Close</Behaviour_Type>
                <Behaviour_Type>Barbarians Not Close</Behaviour_Type>
            </Entity_Behaviours>
        </Entity_Type>
    </Entitiy_Types>
    <Behaviours>
        <Behaviour Type="Leave Settlement">
```

```xml
                    <Ability>1.0</Ability>
                    <Flexibility>1.0</Flexibility>
                    <Collective>
                        <Coordination>1.0</Coordination>
                        <Cooperation>1.0</Cooperation>
                        <Communication>
                            <Signal_In>1.0</Signal_In>
                            <Signal_Out>1.0</Signal_Out>
                        </Communication>
                    </Collective>
                    <Requires>
                        <Behaviour_Type AndOr="Or" PosNeg="Positive">Barbarians Not Close</Behaviour_Type>
                        <Behaviour_Type AndOr="Or" PosNeg="Negative">Barbarians Close</Behaviour_Type>
                    </Requires>
                </Behaviour>
                <Behaviour Type="Barbarians Not Close">
                    <Ability>0.5</Ability>
                    <Flexibility>0.5</Flexibility>
                    <Collective>
                        <Coordination>1.0</Coordination>
                        <Cooperation>1.0</Cooperation>
                        <Communication>
                            <Signal_In>1.0</Signal_In>
                            <Signal_Out>1.0</Signal_Out>
                        </Communication>
                    </Collective>
                </Behaviour>
                <Behaviour Type="Barbarians Close">
                    <Ability>1.0</Ability>
                    <Flexibility>1.0</Flexibility>
                    <Collective>
                        <Coordination>1.0</Coordination>
                        <Cooperation>1.0</Cooperation>
                        <Communication>
                            <Signal_In>1.0</Signal_In>
                            <Signal_Out>1.0</Signal_Out>
                        </Communication>
                    </Collective>
                </Behaviour>
                <Behaviour Type="Move About Countryside">
                    <Ability>1.0</Ability>
                    <Flexibility>1.0</Flexibility>
                    <Collective>
                        <Coordination>1.0</Coordination>
                        <Cooperation>1.0</Cooperation>
                        <Communication>
                            <Signal_In>1.0</Signal_In>
                            <Signal_Out>1.0</Signal_Out>
                        </Communication>
                    </Collective>
                </Behaviour>
            </Behaviours>
        </Problem_Spec>
```